\documentclass{rmf-d}
\usepackage {nopageno, rmfbib, multicol,times,epsf,amsmath,amssymb,cite}
\usepackage[latin1]{inputenc}
\usepackage[]{caption2}
\usepackage{graphicx}
\usepackage{hyperref}
\usepackage{comment}
\usepackage{tabularx}
\usepackage{float}
\usepackage{bbm}
\usepackage{epsfig}

\def\rmfcornisa{Research OR Education of Physics \hfill\rmf %\ {\bf ??} (*?*) ???--???
\hfill  01/2022}

\begin{document}
\title{  Dressed quark-gluon vertex form factors from gauge symmetry
\vspace{-6pt}    }
\author{ Bruno El-Bennich$^1$, Fernando E. Serna$^{1,2}$ and Roberto Correa da Silveira$^1$}
\address{$^1$LFTC, Universidade Cidade de S\~ao Paulo, Rua Galv\~ao Bueno 868, 01506-000 S\~ao Paulo, SP, Brazil \\
$^2$Departamento de F\'isica, Universidad de Sucre, Carrera 28 No. 5-267, Barrio Puerta Roja, Sincelejo, Colombia}
\author{Luis Albino$^3$ and Adnan Bashir$^3$  }
\address{$^3$Instituto de F\'isica y Matem\'aticas, Universidad Michoacana de San Nicol\'as de Hidalgo, Morelia, Michoac\'an 58040, M\'exico }
\author{Eduardo Rojas$^4$}
\address{$^4$Departamento de F\'isica, Universidad de Nari\~no, A.A. 1175, San Juan de Pasto, Colombia}

%%%%%%%%%%%%%%%%%%%%%%%%%%%%%%%%%%%%%%%%%%%%%%%%%%%%%%%%%%%%%%%%%%%%%%%%%%%%%%%%%%%%%%%%%

\maketitle

\recibido{day month year}{day month year}
\vspace{-12pt}
\begin{abstract}
\vspace{1em}  We present preliminary results on the longitudinal \emph{and} transverse form factors of the quark-gluon vertex as functions of the incoming and outgoing quark 
momenta and an angle $\theta =2\pi/3$ between them. The expressions for these form factors were previously derived from Slavnov-Taylor identities, 
gauge covariance and multiplicative renormalizability that firmly constrain the fermion-boson vertex.  \vspace{1em}
\end{abstract}
\keys{ Quantum Chromodynamics, Strong Interactions, Nonperturbative Techniques, Gap Equation, Quark-Gluon Vertex, Dynamical Chiral Symmetry Breaking    \vspace*{-5mm}}
\pacs{  \bf{\textit{12.38.-t	% Quantum chromodynamics
                           11.15.Tk   %	Other nonperturbative techniques
                           02.30.Rz	 % Integral equations
                           11.30.Qc	 % Spontaneous and radiative symmetry breaking
                           14.65.Bt	 % Light quarks
                            }}      \vspace{-4pt}}

%%%%%%%%%%%%%%%%%%%%%%%%%%%%%%%%%%%%%%%%%%%%%%%%%%%%%%%%%%%%%%%%%%%%%%%%%%%%%%%%%%%%%%%%%

\begin{multicols}{2}

The origin of \emph{dynamical chiral symmetry breaking} (DCSB), the mass-generating mechanism responsible for the overwhelming contribution to the nuclei's masses, 
lies in the non-Abelian nature of the theory of strong interactions known as Quantum Chromodynamics (QCD). Starting with the seminal work by Nambu and 
Jona-Lasinio~\cite{Nambu:1961tp}, this mechanism has been gradually elucidated in QCD over the past decades. While its role in generating hadron masses two 
orders of magnitude larger than those of the light current quarks is nowadays widely recognized, its likely deeper connection to \emph{confinement} 
still remains speculative.  

A common approach to investigate DCSB is to study the gap equation of the quark, i.e. its two-point Green function, and its nonperturbative formulation in terms of the 
Dyson-Schwinger equation (DSE)~\cite{Bashir:2012fs}. The latter is a Euclidean-space description of the quark's equation of motion in relativistic quantum field theory 
and can be derived from the generating functional in QCD~\cite{Roberts:1994dr}. The self-energy term in this integral equation involves other Green functions, namely 
the gluon propagator and the quark-gluon  vertex, which  are irreducible two- and three-point functions, respectively. Both contribute, along with the strong coupling 
$\alpha_s$, to the integral kernel's ``strength" in  the DSE. 

Indeed, it is this strength that controls the emergence of a mass gap. While the strong coupling and the form factor associated with the gluon dressing form an overall 
strength  factor, the twelve tensor structures of the quark-gluon vertex reveal  a more intricate story. Their contributions are codified in  so-called longitudinal and transverse
form factors and their convolution with the strong coupling and the gluon dressing function gives rise to a constituent quark mass scale.

Since the full structure of the dressed quark-gluon vertex, and in particular of the associated form factors, still poses a serious  computational challenge in functional as well as 
in lattice-QCD approaches,  a common expedient in applications to hadron physics is to retain merely its perturbative $\gamma_\mu$ term. Folding its form factor with that of 
the gluon propagator and the strong coupling, one arrives at the nowadays well known \emph{rainbow-ladder} truncation of the gap equation, in which a single analytic function 
mimics the infrared and ultraviolet behavior and the strength of the strong interaction in an effective manner~\cite{Maris:1999nt,Qin:2011dd}. 

While this approach has certainly proven to be successful in the computation of the light meson and baryon spectrum and their electromagnetic properties 
\cite{Rojas:2014aka,Raya:2015gva,El-Bennich:2016qmb,Mojica:2017tvh,Cloet:2008re,Eichmann:2009qa,Segovia:2015hra,Chen:2017pse,Eichmann:2016hgl,
Eichmann:2016yit}, it fails to correctly describe the scalar and axialvector meson masses and does not produce a satisfying mass ordering of higher radially excited 
mesons.  It also gives rise to  a spurious spectrum of unobserved light mesons~\cite{Rojas:2014aka} with ``exotic" quantum numbers, and admits $\bar 3_c$ colored 
diquark bound-states.  Still, the latter feature  can be favorably used to derive approximate Faddeev wave functions of baryons. The shortcoming of this leading approximation 
is also  observed in solving the Bethe-Salpeter equation for pseudoscalar and vector $D$ and $B$ mesons~\cite{Rojas:2014aka,Mojica:2017tvh}, but can be overcome 
introducing  a flavor dependence in the quark-gluon  interaction~\cite{Serna:2017nlr,Serna:2020txe,El-Bennich:2021ldv}. However, this comes at the cost of additional 
parameters for the charm and bottom mesons. 

Important improvements, based on the three-particle irreducible (3PI) effective QCD action~\cite{Williams:2015cvx} or on a model ansatz for the quark-gluon 
vertex~\cite{Chang:2011ei,Binosi:2016wcx} amongst others, have been obtained over the past decade, and one may assert that functional QCD approaches to light 
and flavored mesons, heavy quarkonia and baryons based on the DSE in conjunction with either the Bethe-Salpeter equation (BSE) or Faddeev equation are overall 
very successful. This includes the mass spectrum of light and heavy mesons, the nucleon and $\Delta$ baryons, their parity partners and radial excitations, as well 
as Compton scattering, elastic and transition form factors. Extensions to tetraquark states have also been studied within this approach~\cite{Heupel:2012ua}. 

Nonetheless, a less model dependent interaction kernel, based on \emph{calculated\/} QCD Green functions, of the quark DSE and related bound-state equations is desirable. 
Only a detailed construction of the interaction kernel, which involves the fully dressed  quark-gluon vertex, will allow to verify whether the known hadron spectrum can be 
completely described with functional methods in QCD. In addition, there are not merely phenomenological but also formal, field-theoretical motivations to study 
the analytic behavior of the fermion-boson vertex. After all, this vertex plays a pivotal role for DCSB in QED and in QCD. Its contribution to the infrared behavior of 
the quark propagator and to fragmentation functions, and therefore to the elucidation of the confinement mechanism, cannot be appreciated enough. 

In this contribution we extend recent studies on the transverse quark-gluon vertex, which we derived from transverse Slavnov-Taylor identities and multiplicative 
renormalizability in Refs.~\cite{Serna:2018dwk,Albino:2021rvj}. In those studies we did not present figures of the different vertex form factors, so we 
here take the opportunity to fill this gap.

%%%%%%%%%%%%%%%%%%%%%%%%%%%%%%%%%%%%%%%%%%%%%%%%%%%%%%%%%%%%%%%%%%%%%%%%%%%%%%%%%%%%%%%%%

The dressed quark-gluon vertex is the essential three-point function which describes the nonperturbative coupling of a dressed quark to a dressed gluon. As such, 
it is the source of nonperturbative radiative gluon corrections to the current quark's relativistic motion. As in other relativistic quantum field theories, the related equation of motion
can be expressed by a DSE conveniently derived within a functional approach to QCD~\cite{Roberts:1994dr}:
\begin{align}
\hspace*{-2mm}
     S^{-1}(p) & =  \, Z_2 \,  i \gamma\cdot p  + Z_4 \, m  \nonumber \\
    & \, + Z_1 g^2\!  \int^\Lambda\!\!  \frac{d^4k}{(2\pi)^4}\  \Delta^{ab}_{\mu\nu} (q)\, \gamma_\mu t^a  S(k) \, \Gamma_\nu^b (k,p) \ . 
 \label{DSEquark}
\end{align}
In this integral equation, $m$ is the renormalized current-quark mass and $Z_4(\mu,\Lambda)$ is its renormalization constant in the QCD Lagrangian, while  $Z_1(\mu,\Lambda)$ 
and $Z_2(\mu,\Lambda)$ are vertex and wave-function renormalization constants, respectively. The integral in Eq.~\eqref{DSEquark} expresses  the quark's self-energy 
$\Sigma (p^2)$,  where $\Lambda$ is an ultraviolet Poincar\'e invariant cut-off and $\mu$ is the renormalization scale imposed, so that $S^{-1}(p)|_{p^2=\mu^2} = \gamma\cdot p 
+ m(\mu)$, with the common choice $\Lambda \gg \mu$. Furthermore, $\Delta_{\mu\nu}(q)$ is the dressed gluon propagator in Landau gauge and $\Gamma^a_\mu (k,p) 
= \Gamma_\mu (k,p)\, t^a $ is the dressed quark-gluon vertex, where $t^a = \lambda^a/2$  are the SU(3)$_c$ group generators and $a,b$ represent color indices.

The solutions of the DSE~\eqref{DSEquark} can be most generally decomposed into scalar and vector pieces,
\begin{equation}
\label{DEsol}
   S (p)  =   \frac{1}{i \gamma \cdot p \,A (p^2)   + B ( p^2 ) } = \,  \frac{Z (p^2 )}{ i \gamma \cdot p + M ( p^2 )} \   ,
\end{equation}
where $Z(p^2,\Lambda^2,\mu^2) = 1/A(p^2,\Lambda^2,\mu^2)$ and $M (p^2) = B(p^{2},\mu^2,\Lambda^2)/A(p^{2},\mu^2,\Lambda^2)$ are the flavor dependent, running 
wave renormalization and mass functions, respectively. 

The complete vertex, $\Gamma_\mu (k,p)$, can be expanded in terms of four non-transverse and eight transverse covariant vector structures~\cite{Ball:1980ay},
\begin{align}
\hspace*{-2mm}
    \Gamma_{\mu}(k,p)  & = \, \Gamma^L_{\mu}(k,p) +\Gamma^T_{\mu}(k,p)  \nonumber \\
     = &\ \sum_{i=1}^{4}\, \lambda_{i}(k,p) L^i_\mu (k,p)  + \sum_{i=1}^{8} \tau_{i}(k,p) T^i_\mu (k,p)\,  ,
\label{BallChiu}
\end{align}
in which $p$ is the incoming and $k$ the outgoing quark momentum and the gluon momentum, $q=k-p$, flows into the vertex. The transverse vertex is naturally defined 
by  $ q \cdot \Gamma^T (k,p)  =  0$.  We work with the vector base for $L^i_\mu (k,p)$ and $T^i_\mu (k,p)$ defined in Ref.~\cite{Kizilersu:1995iz}.  
 
The form factors of the quark-gluon vertex, in particular of the longitudinal components, have been explored in pQCD and in nonperturbative approaches, see for instance 
Refs.~\cite{Alkofer:2008tt,Williams:2014iea,Bashir:2011dp,Bermudez:2017bpx,Rojas:2013tza,Aguilar:2016lbe,Aguilar:2018epe,
Binosi:2016wcx,Oliveira:2020yac,Pelaez:2015tba,Skullerud:2003qu,Kizilersu:2021jen} or Ref.~\cite{Albino:2021rvj} for a more detailed  bibliography on the fermion-boson 
vertex. Recently, we employed  two transverse Slavnov-Taylor identities~\cite{He:2009sj}, which express color gauge invariance and Lorentz covariance and constrain
 the transverse quark-gluon vertex,  to derive the  eight $\tau_{i}(k,p)$ form factors in  QCD~\cite{Albino:2021rvj}. Along with the known expressions for $\lambda_{i}(k,p)$
\cite{Rojas:2013tza,Aguilar:2016lbe} we found that $ \Gamma_{\mu}(k,p)$ in Eq.~\eqref{BallChiu} can be described by the following set of form factors:
\begin{eqnarray}
  \lambda_1  (k, p)  & = & \tfrac{1}{2}\,  G(q^2) X_0   (q^2)\left[ A  (k^2) + A  (p^2)   \right] \ , 
 \label{lambda1QCD}   \\  [3pt] 
   \lambda_2  (k, p)  & = &  G(q^2) X_0   (q^2)\, \frac{ A   (k^2) - A  (p^2)}{k^2-p^2}  \ , 
 \label{lambda2QCD}   \\  [3pt] 
  \lambda_3  (k, p)  & = &   G(q^2) X_0   (q^2)\, \frac{ B   (k^2) - B  (p^2)}{k^2-p^2}  \ , 
 \label{lambda3QCD}   \\  [3pt] 
  \lambda_4  (k, p)  & = &\ 0 \ ,  \\  [3pt] 
 \label{lambda4QCD}  
   \tau_1  (k, p) & = & - \frac{ Y_{1} }{ 2 (k^{2} - p^{2}) \nabla(k,p) } \ ,
\label{tau1QCD}   \\  [3pt]   
   \tau_2  (k, p) &=& - \frac{Y_{5} - 3 Y_{3}}{ 4 (k^{2} - p^{2}) \nabla(k,p) } \ ,
\label{tau2QCD} \\ [3pt] 
   \tau_3  (k, p) &=&  \frac{1}{2}\,  G(q^2) X_0   (q^2)\left[ \frac{ A   (k^2) - A  (p^2)}{k^2-p^2}  \right] \nonumber \\
   & + & \frac{Y_{2}}{4\nabla(k,p)}  - \frac{ (k+p)^{2} (Y_{3} - Y_{5}) }{ 8(k^{2} - p^{2}) \nabla(k,p) } \ ,   
\label{tau3QCD}   \\  [3pt] 
   \tau_4  (k, p) &=&  - \frac{ 6 Y_{4} + Y_{6}^{A} }{8\nabla(k,p) } - \frac{(k+p)^{2} Y_{7}^{S} }{ 8(k^{2} - p^{2}) \nabla(k,p) } \  , 
\label{tau4QCD}  \hspace*{5mm}  \\ [3pt] 
 \tau_5  (k, p) &=& -  G(q^2) X_0   (q^2)\left[ \frac{ B   (k^2) - B  (p^2)}{k^2-p^2}  \right ]   \nonumber   \\
      & - &  \frac{2 Y_4 + Y_6^A}{2 (k^{2}-p^{2})} \ , 
\label{tau5QCD}    \\ [3pt] 
 \tau_6  (k, p) &=& \frac{(k-p)^{2} Y_{2} }{ 4 (k^{2} - p^{2}) \nabla(k,p) } -\frac{Y_3 - Y_5}{8\nabla(k,p) }  \ , 
\label{tau6QCD} %\\ [3pt] 
\end{eqnarray}
%
%%%%%%%%%%%%%%%%%%%%%%%%%%%%%%%%%%%%%%%%%%%%%%%%%%%%%%%%%%%%%%%%%%%%%%%%%%%%%%%%%%%%%%%%
%
\begin{figure*}[t!]
\vspace*{-5mm}
\centering
   \includegraphics[scale=0.5]{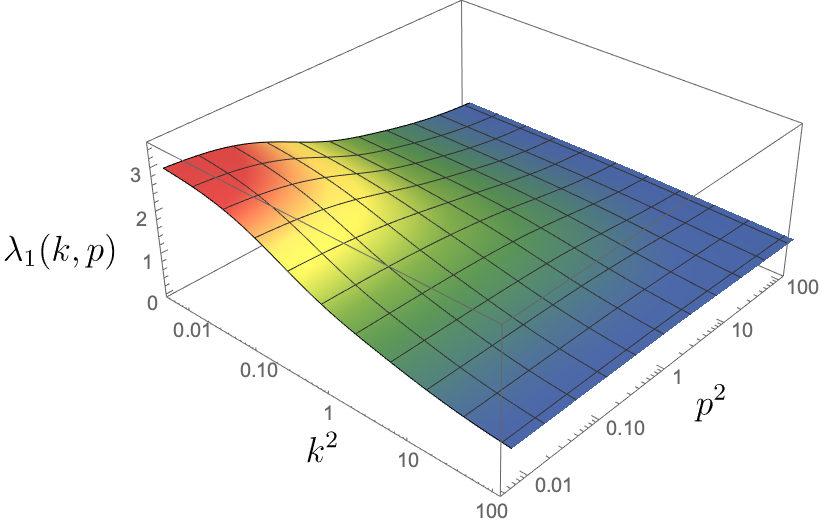}    \hspace*{1cm}
   \includegraphics[scale=0.5]{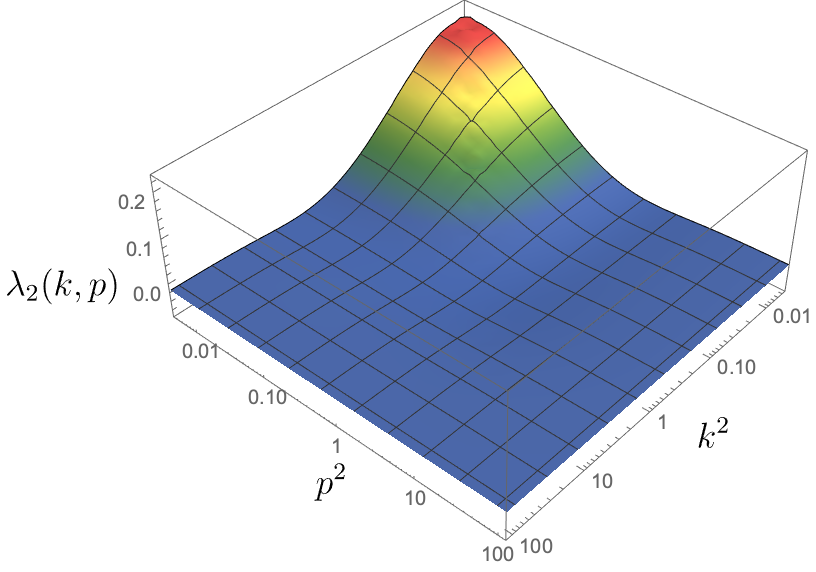}   
   \includegraphics[scale=0.5]{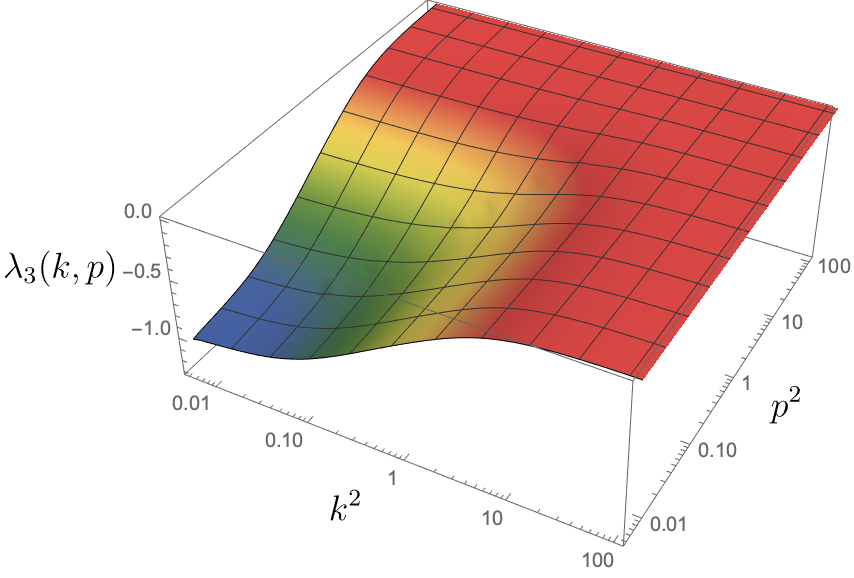}  \\  \vspace*{2mm} 
\caption{Form factors of the longitudinal vertex $\Gamma^L_{\mu}(k,p)$~\eqref{BallChiu}  as functions of the quark momenta, $k^2$ and $p^2$ [in GeV$^2$], 
and for the angle $\theta= 2\pi/3$.   \vspace*{-2mm} }  
\label{fig1}
\end{figure*}
%%%%%%%%%%%%%%%%%%%%%%%%%%%%%%%%%%%%%%%%%%%%%%%%%%%%%%%%%%%%%%%%%%%%%%%%%%%%%%%%%%%%%%%%% 
%
\vspace*{-4mm}
\begin{eqnarray}
   \tau_7  (k, p) &=& \frac{q^2 (6Y_4 +Y_6^A)}{4(k^2-p^2)\nabla(k,p)}   +\frac{Y_7^S}{4\nabla(k,p )} \ , 
\label{tau7QCD} \\ [3pt] 
   \tau_8  (k, p) &=&  - G(q^2) X_0   (q^2)\left[ \frac{ A   (k^2) - A  (p^2)}{k^2-p^2}  \right]  \nonumber \\
   & - &  \frac{2 Y_8^A}{k^{2}-p^{2}}  \ .
\label{tau8QCD}
\end{eqnarray}
%
%%%%%%%%%%%%%%%%%%%%%%%%%%%%%%%%%%%%%%%%%%%%%%%%%%%%%%%%%%%%%%%%%%%%%%%%%%%%%%%%%%%%%%%%

\begin{figure*}[t!]
\vspace*{-5mm}
\centering
   \includegraphics[scale=0.43]{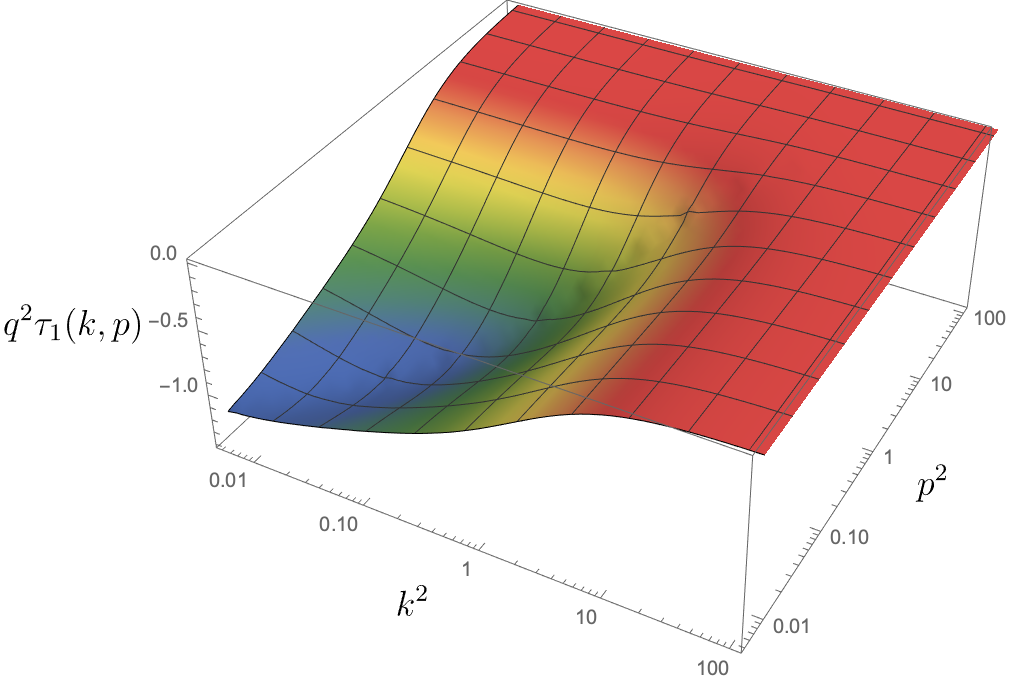}    \hspace*{1.2cm}
   \includegraphics[scale=0.44]{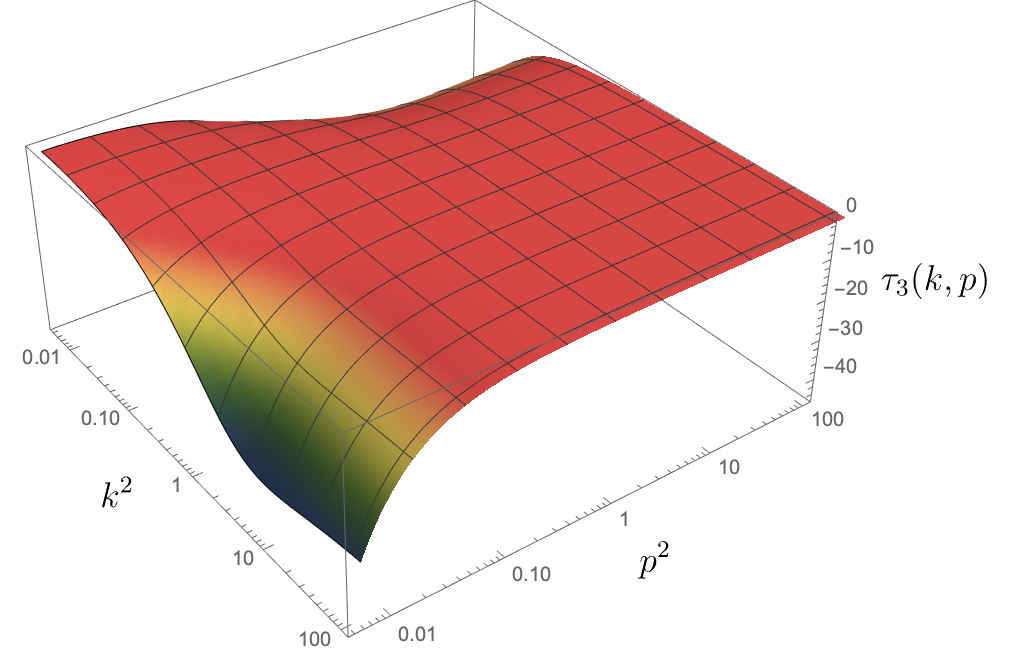}   \\
   \includegraphics[scale=0.47]{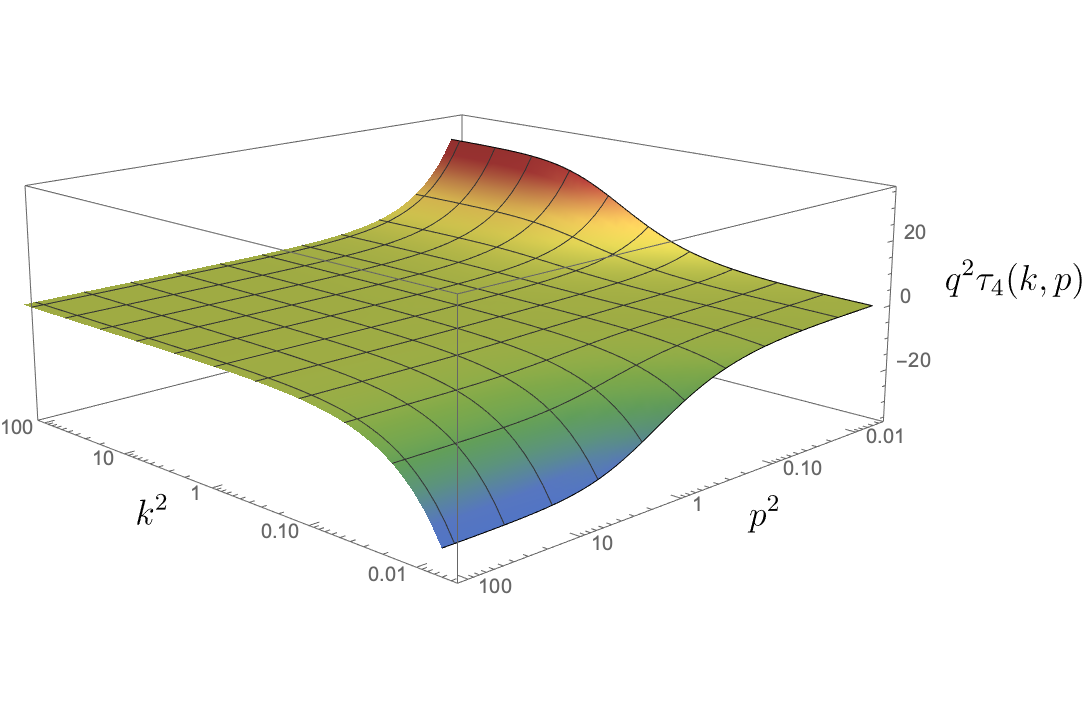}     \hspace*{3mm}
   \includegraphics[scale=0.48]{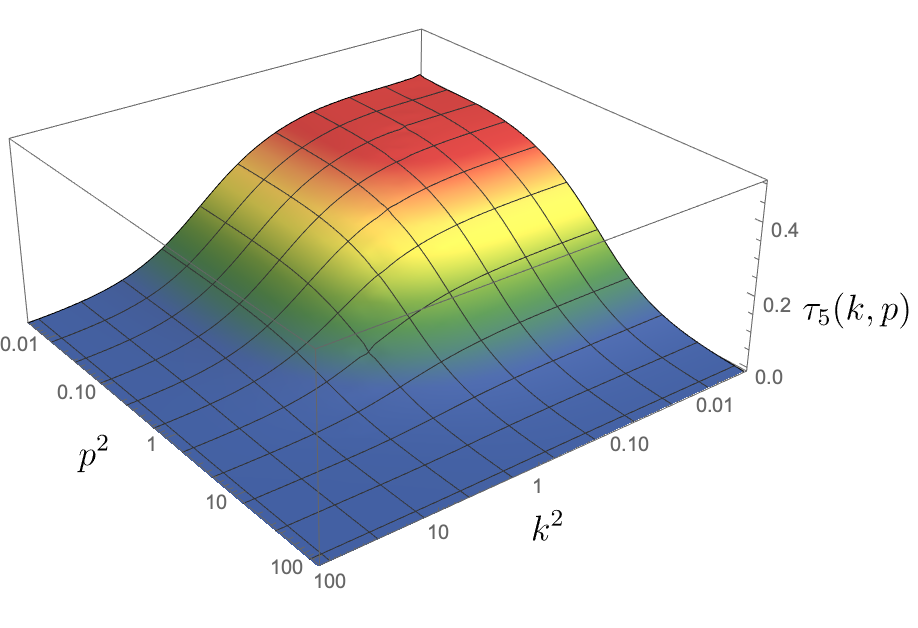}     \\
   \includegraphics[scale=0.46]{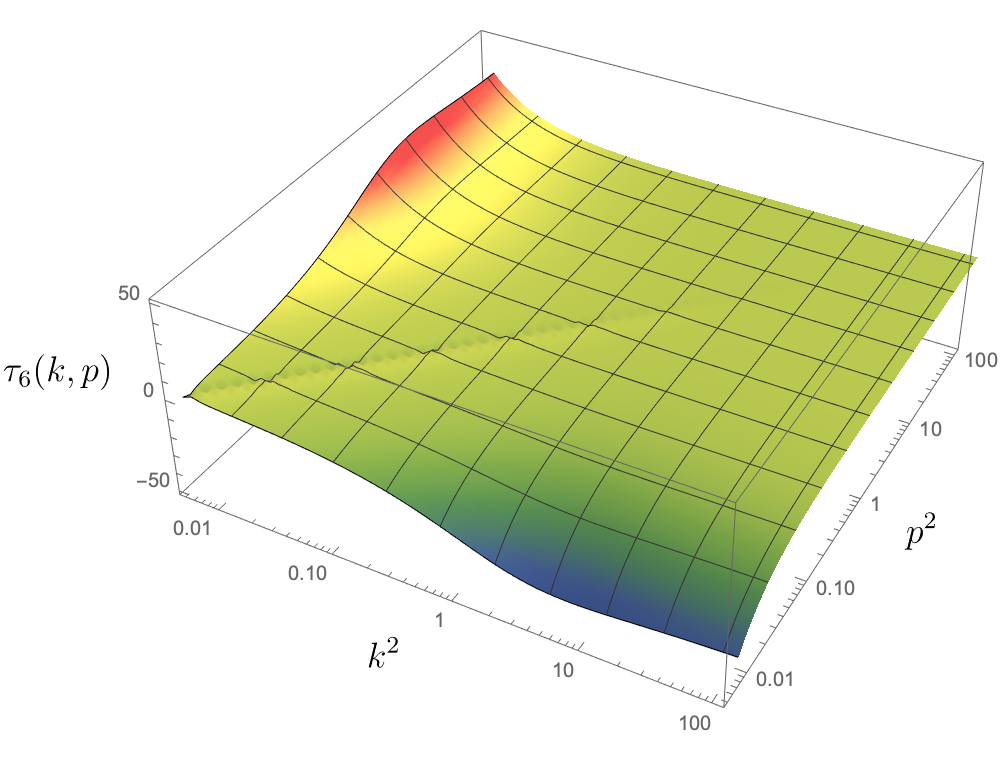}     \hspace*{1.3cm} 
   \includegraphics[scale=0.47]{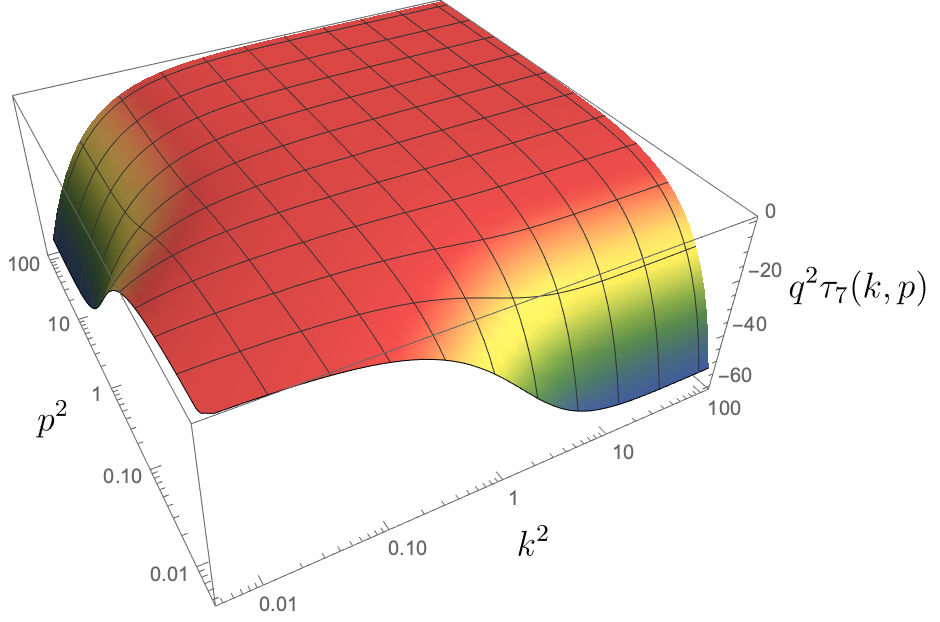}   \\ \vspace*{2mm}
\caption{Form factors of the transverse vertex $\Gamma^T_{\mu}(k,p)$~\eqref{BallChiu} as functions of the quark momenta, $k^2$ and $p^2$ [in GeV$^2$], and 
for the angle $\theta= 2\pi/3$.}
\label{fig2}
\end{figure*}

%%%%%%%%%%%%%%%%%%%%%%%%%%%%%%%%%%%%%%%%%%%%%%%%%%%%%%%%%%%%%%%%%%%%%%%%%%%%%%%%%%%%%%%%% 

In Eqs.~\eqref{tau1QCD} to \eqref{tau8QCD} the Gram determinant is defined by $ \nabla (k,p) = k^{2} p^{2} - (k \cdot p)^{2} $. The form factors $\lambda_i (k, p)$, 
$i=1,2,3$, and $\tau_i (k, p)$, $i=3,5,8$, are proportional to the ghost-dressing function $G(q^2)$ which is renormalized as  $G(\mu^2) = 1 $. Moreover, $X_0 (q^2)$ is 
the leading form factor of the quark-ghost scattering amplitude, $H^{a} (k, p) = H (k, p) t^a$; see, e.g., Refs.~\cite{Aguilar:2016lbe,Aguilar:2018epe} for details.  
The a priori unknown scalar functions, $Y_i^{A,S}$, are form factors we introduce to decompose a four-point function that appears in
the transverse  Slavnov-Taylor identities and which involves a non-local vector vertex and a Wilson line to preserve gauge invariance. We refer to the discussion
in Ref.~\cite{He:2009sj} and merely stress that the $Y_i^{A,S}$ functions  have been constrained by us in Ref.~\cite{Albino:2018ncl} with the vertex ansatz of
Ref.~\cite{Bashir:2011dp} and insisting on multiplicative renormalizability.

We solve the DSE~\eqref{DSEquark} with numerical input from lattice QCD~\cite{Dudal:2018cli,Duarte:2016iko}  for the gluon and ghost dressing functions, $\Delta (q^2)$
and $G(q^2$), and with the vertex defined by Eqs.~\eqref{BallChiu} to \eqref{tau8QCD}. The solution for the vector and scalar components of the quark propagator, $A (p^2)$ 
and $B (p^2)$ respectively, are then used to compute the form factors $\lambda_{i}(k,p)$ and  $\tau_{i}(k,p)$. We present them in Figs.~\ref{fig1} and \ref{fig2} as 
functions of the momenta squared $k^2$ and $p^2$ and for the kinematic configuration: $\cos \theta = k\cdot p / |k||p| = -1/2 $, $\theta =2\pi/3$. 

The functional form of the $\lambda_i (k, p)$  form factors is similar to that found in Refs.~\cite{Aguilar:2016lbe,Aguilar:2018epe} though with differences in magnitude, as those
studies exclusively concentrated on $\Gamma^L_{\mu}(k,p)$. Therefore, some strength is shifted from the transverse to the longitudinal vertex and a direct comparison is difficult. 
As observed  in  Ref.~\cite{Albino:2021rvj}, the dominating contribution of the transverse vertex in the gap equation~\eqref{DSEquark}, and therefore to DCSB, is due to the form 
factors  $\tau_4(k, p)$ and $\tau_7(k, p)$. We multiply them and also  $\tau_1 (k, p)$ by $q^2$ in Fig.~\ref{fig2} in order to regularize a singular behavior at the origin. Note that this 
poses no problem in the numerical treatment of the quark DSE since kinematic factors in the integral kernel  have an analogous regularizing effect.  

A more comprehensive treatment of the these form factors taking into account other angles $\theta$, which include the soft-gluon and symmetric quark limit, 
is underway. This also requires the contributions of the sub-leading form factors $X_1(k,p)$, $X_2(k,p)$ and $X_3(k,p)$ that parametrize the quark-ghost scattering 
amplitude. \bigskip

\noindent
\textbf{Acknowledgments}

\noindent
B.E.'s research activity is funded by FAPESP grant no.~2018/20218-4 and CNPq grant no.~428003/2018-4. F.E.S. is a CAPES-PNPD postdoctoral fellow, grant no.~88882.314890/2013-01,
L.A. was a FAPESP postdoctoral fellow supported by grant no.~2018/17643-5, and R.C.S. is a CAPES PhD fellow. A.B. acknowledges funding from the Coordinaci\'on de la Investigaci\'on 
Cient\'ifica (CIC) of the University of Michoac\'an and CONACyT, Mexico, through  grant nos.~4.10 and CB2014-22117, respectively.

%%%%%%%%%%%%%%%%%%%%%%%%%%%%%%%%%%%%%%%%%%%%%%%%%%%%%%%%%%%%%the%%%%%%%%%%%%%%%%%%%%%%%%%%%%

\end{multicols}
\medline 

%%%%%%%%%%%%%%%%%%%%%%%%%%%%%%%%%%%%%%%%%%%%%%%%%%%%%%%%%%%%%%%%%%%%%%%%%%%%%%%%%%%%%%%%%

\begin{multicols}{2}

\end{multicols}
\end{document}